\documentclass[12pt,dvips]{article}
\usepackage{graphicx}

\newcommand{\I}{\mathrm{i}}
\newcommand{\MSbar}{\ensuremath{\overline{\textrm{MS}}}}

\newcommand{\dbar}{\ensuremath{\overline{d}}}
\newcommand{\qbar}{\ensuremath{\overline{q}}}
\newcommand{\be}{\begin{equation}}
\newcommand{\ee}{\end{equation}}
\newcommand{\eq}[1]{(\ref{#1})}

\begin{document}

\thispagestyle{empty}

\date{\today}
\title{
\vspace*{2.5cm}
The  chiral and angular momentum content \\of the
$\rho$-meson\footnote{This paper is dedicated to our colleague Willibald Plessas
on occasion of his 60th birthday and will appear in special issue of {\em Few Body
Systems}.}}
\author{L. Ya. Glozman, C. B. Lang, and M. Limmer\\~\\
Institut f\"ur Physik, FB Theoretische Physik, \\
Universit\"at Graz, A-8010 Graz, Austria}
\maketitle

\begin{abstract}
It is possible to define and calculate in a gauge-invariant manner the chiral
as well as the partial wave content of the  quark-antiquark Fock component of
a meson in the infrared, where mass is generated. Using the variational method
and a set of interpolators that span a complete chiral basis we extract in a
lattice QCD Monte Carlo simulation with $n_f = 2$ dynamical light quarks the
orbital angular momentum and spin content of the $\rho$-meson. We obtain in
the infrared a simple $^3S_1$ component as a  leading component of the
$\rho$-meson with a small admixture of the $^3D_1$ partial wave, in agreement
with the $SU(6)$ flavor-spin symmetry.
\end{abstract}


\newpage

\section{Introduction}

The $SU(6)_{FS}$ flavor-spin symmetry for the low-lying hadrons in the light
flavor sector \cite{SU6} and its roots (as due to the nonrelativistic quark
model \cite{NQM}) predated QCD and had numerous phenomenological successes.
When QCD was established as the fundamental theory of strong interactions it
has soon become clear that the current quarks of QCD in the $u,d$ sector have
tiny masses, of the order of a few MeV at the renormalization scale of 1-2
GeV. They are very far away from the rather heavy constituent quarks of the quark
model. In view of this the QCD Lagrangian has approximate chiral symmetry. 

At the same time it was clear that this approximate chiral symmetry is
dynamically broken in the vacuum and this breaking is a source of mass of the
low-lying hadrons. Different microscopical models (with varying definitions of
the quark mass) exist for chiral symmetry breaking in the vacuum and they
indicate that indeed at large space-like momenta the quark mass matches its
bare values of a few MeV, while at low momenta it runs to the value of the
constituent quark mass in the definitions given by the models. There also
exist lattice determinations of such masses at low momenta, see, e.g.,
\cite{Aoki,Para}, though such determinations are manifestly gauge dependent.
By fixing a gauge and considering a single quark propagator in the background
gluonic field it is not clear a priori how the confining gluodynamics (that
drives the dynamics in the color-singlet hadrons) is taken into account. Then
it is interesting to see how it would be possible to provide a bridge from QCD
in the infrared to the language of the quark model in a model-independent and
gauge-invariant manner.

There is a systematic method to study the hadron composition on the lattice
- the variational method \cite{Mi85,LuWo90}. In that approach one selects a
set of interpolating operators $\{O_1,\,O_2,\, ...,\, O_N\}$ with the proper
quantum numbers that couple to a given hadron. One computes the
cross-correlation matrix
\begin{equation}\label{eq_corrfctn}
C_{ij}(t) = \langle O_i(t) O_j^\dagger(0) \rangle.
\end{equation}
Masses of the ground and excited states of hadrons with  fixed quantum numbers
can be extracted from the  $t$-dependence of the eigenvalues of  this matrix at
large Euclidean times $t$. If the set of operators $O_i$ is complete enough,
then the eigenvectors represent the ``wave function'' of the hadron (see  the
cautionary remarks at the begin of Sect.\ \ref{sec_var_meth} and the definition
\eq{eq_w_f}). Of course, hadrons contain many different Fock components. Our
task is to reconstruct the leading one, the quark-antiquark component of the
low-lying mesons. For this one needs  a set of operators that allows one to
define uniquely such a component. This set of operators must be complete in the
$\qbar q$ space with regard  to the chiral basis.

All possible $\qbar q$ interpolators for non-exotic mesons in the $u,d$ sector
have been classified according to transformation properties with respect to
$SU(2)_L \times SU(2)_R$ and $U(1)_A$ chiral groups \cite{Gl03,Gl07}. If one
assumes that there is no explicit excitation of the gluonic field with the
non-vacuum quantum numbers, which is certainly true for low-lying hadrons,
then the $SU(2)_L \times SU(2)_R$ representations for the quark-antiquark
system specify a complete and orthogonal basis. Consequently a set of
interpolators that is in one-to-one correspondence with  all possible chiral
representations of  $SU(2)_L \times SU(2)_R$ is a complete one and can be used
to define the $\qbar q$ component of a meson. The cross-correlation matrix
with such a  set can be used to reconstruct the $\qbar q$ Fock component.  The
eigenvectors of this correlation matrix represent the $\qbar q$ content
in terms of different chiral representations. Observing a
superposition of different chiral representations implies that chiral
symmetry is broken.

It turns out that it is also possible to reconstruct a composition of the
$\qbar q$ component in terms of the $^{2S+1}L_J$ basis, where {\bf J = L + S}
are standard angular momenta. Indeed, the complete and orthogonal chiral basis
can be related, through a unitary transformation, to the complete and
orthogonal $^{2S+1}L_J$ basis in the center-of-momentum frame \cite{GlNe07}.
Then diagonalizing the cross-correlation matrix with interpolators that span a
complete set of chiral representations and using this unitary transformation
to the $^{2S+1}L_J$ basis one can obtain from the eigenvectors of the
correlation matrix a partial wave decomposition of the $\qbar q$ component.

In QCD the decomposition of a hadron should depend on the scale at which we
probe this hadron. In other words, what we see in our microscope depends on
the resolution. In our case ``the microscope'' is our interpolating operator
$O_i$ that creates the hadron from the vacuum. A true point-like source would
correspond to the point-like lattice interpolator in the limit of the lattice
spacing approaching 0, $a \rightarrow 0$. The point-like interpolator applied
on the lattice with the spacing $a$ probes the hadron at the scale specified
by $a$. In the continuum limit it becomes the true point-like operator and
probes the hadron at the scale $\mu^2 \rightarrow \infty$. Here we want to
study the hadron structure at the infrared scale, where the mass is generated.
This scale is determined by the hadron size, of the order 0.3 - 1 fm. In
lattice simulations we cannot use such a large $a$, because then the lattice
artifacts are too large and matching to continuum theory is lost. However,
such a low scale can be fixed by the gauge-invariant smearing of the
interpolators.  If we use the interpolating operator smeared over the size $R$
in the physical units such that $R/a \gg 1$, then even in the continuum limit
$a \rightarrow 0$ we probe the hadron structure at the scale $R$. Changing the
smearing size $R$ we can study  the hadron content at different scales of the
continuum theory at $a \rightarrow 0$. 

In this paper we expand our results on the chiral and partial wave
decomposition of the $\rho$-meson in the infrared, presented in a recent
letter \cite{GlLaLi09}, discuss all required details of  the formalism and
physical interpretation as well as give some additional numerical results.

\section{Chiral classification of the quark-antiquark interpolators}

The chiral classification of some of the $\qbar q$ interpolators  was done in
\cite{CJ}. A complete classification was performed in \cite{Gl03,Gl07} and is
summarized here.

We consider  the two-flavor mesons. All quark-antiquark bilinear operators in
the chiral limit can be classified according to the representations of the
$SU(2)_L \times SU(2)_R$ and $U(1)_A$ chiral groups.  Consider, as example,
local interpolators of $J=0$ mesons,  built from quark isodublets $q$: 
\begin{eqnarray}
O_\pi(x)  &=& \I \,\qbar(x)\, \vec \tau\,\gamma_5\, q(x)\;,
\label{ppi}
\\
 O_{f_0}(x)  &=&  \qbar(x)\,  q(x)\;,
\label{ff0}
\\ 
O_{\eta}(x)  &=&  \I \,\qbar(x) \,\gamma_5\, q(x)\;,
\label{eeta}
\\ 
O_{a_0}(x)  &=&  \qbar(x)\, \vec \tau  \,q(x)\;,
\label{aa0}
\end{eqnarray}
where $\vec\tau$ denotes the vector of isospin Pauli matrices. The
$SU(2)_L\times SU(2)_R$ transformations consist of vectorial and axial
transformations in the isospin space. The axial transformation  mixes the
currents of opposite parity:
\begin{equation}
 O_\pi(x)  \leftrightarrow O_{f_0}(x) 
\label{pif0}
\end{equation} 
as well as
\begin{equation}
 O_{a_0}(x)  \leftrightarrow O_{ \eta}(x).
\label{a0eta}
\end{equation} 
Hence the currents (\ref{pif0}) form the basis functions of the $(1/2,1/2)_a$
representation of the chiral  $SU(2)_L\times SU(2)_R$ while the interpolators
(\ref{a0eta}) transform as $(1/2,1/2)_b$.

As another example consider local interpolators of the $\rho$-meson. There exist
two different bilinear operators with the $\rho$-meson quantum numbers that have
radically different chiral transformation properties. The first one is the
standard vector current, 
\begin{equation}
O_\rho^V(x)  = \qbar(x)\, \gamma^i \vec \tau\, q(x)\;,
\label{rV}
\end{equation}
and the second one is the pseudotensor operator,
\begin{equation}
O_\rho^T(x)  = \qbar(x)\, \sigma^{0i} \vec \tau\, q(x).
\label{rT}
\end{equation}

\begin{table}[tb]
\caption[]{The complete set of $\bar{q}q$ states (interpolators)  classified
according to  the chiral basis. The symbol $\leftrightarrow$ indicates the
states (interpolators) belonging to the same representation of $SU(2)_L\times
SU(2)_R$.}\label{table_1}
\begin{center}
\begin{tabular}{cccc}
\hline\noalign{\smallskip}
$R$&$J=0$&$J=1,3,\ldots$&$J=2,4,\ldots$\\
\noalign{\smallskip}\hline\noalign{\smallskip}
$(0,0)$&\bf ---&$0J^{++} \leftrightarrow 0J^{--}$&$0J^{--}  \leftrightarrow 0J^{++}$\\
$(1/2,1/2)_a$&$1J^{-+}\leftrightarrow 0J^{++}$&$1J^{+-} \leftrightarrow 0J^{--}$&$1J^{-+} \leftrightarrow 0J^{++}$\\
$(1/2,1/2)_b$&$1J^{++}\leftrightarrow 0J^{-+}$&$1J^{--} \leftrightarrow 0J^{+-}$&$1J^{++} \leftrightarrow 0J^{-+}$\\
$(0,1) \oplus (1,0)$&\bf ---&$1J^{--} \leftrightarrow 1J^{++}$&$1J^{++} \leftrightarrow 1J^{--}$\\
\noalign{\smallskip}\hline
\end{tabular}
\end{center}
\end{table}

The axial $SU(2)_L\times SU(2)_R$ transformation mixes the vector current
with the axial vector current,
\begin{equation}
O_{a_1}(x)  = \qbar(x)\, \gamma^i \gamma^5 \vec \tau\, q(x),
\label{a1}
\end{equation}
while the pseudotensor interpolator gets mixed with the interpolator of the
$h_1$ meson,
\begin{equation}
O_{h_1}(x)  = \varepsilon ^{ijk} \qbar(x)\, \sigma^{jk} \, q(x).
\label{h1}
\end{equation}
Consequently the operators 
\begin{equation}
 O_\rho^V(x)  \leftrightarrow O_{a_1}(x) 
\label{vaV}
\end{equation} 
form a basis of the $(0,1)\oplus(1,0)$ representation of $SU(2)_L\times SU(2)_R$,
while the interpolators
\begin{equation}
 O_\rho^T(x)  \leftrightarrow O_{h_1}(x) 
\label{vaT}
\end{equation} 
transform as $(1/2,1/2)_b$. A complete set of representations of
$SU(2)_L\times SU(2)_R$ and the corresponding quantum numbers are listed in
Table \ref{table_1}.

In Table \ref{table_1} the index $R$ determines a representation of the
$SU(2)_L \times SU(2)_R$ with $R=$ $(0,0)$, $(1/2,1/2)_a$, $(1/2,1/2)_b$, or
$(0,1)\oplus(1,0)$. All  states (interpolators) are uniquely specified by the
set of quantum numbers $\{R;IJ^{PC}\}$ where we use the standard notations for
the isospin $I$, total spin $J$ as well as for the spatial and charge parities
$PC$. The chiral basis  $\{R;IJ^{PC}\}$ is obviously  consistent with
Poincar{\'e} invariance. The symbol $\leftrightarrow$ indicates that both
given states (interpolators) are members of a particular chiral multiplet,
that is they transform into each other upon $SU(2)_L \times SU(2)_R$.

For a particle with the $\rho$-meson quantum numbers the  $(0,1)\oplus(1,0)$
and the  $(1/2,1/2)_b$ representations are a complete set.  Both
interpolators, $O_\rho^V$ and $O_\rho^T$, may create the $\rho$-meson from the
vacuum. If both couple, indeed, this signals chiral symmetry breaking.

\section{Transformation from the chiral to the angular momentum basis}

The chiral basis can be related, through a unitary transformation, to the 
$\{I;{}^{2S+1}L_J\}$ basis  in the center-of-momentum frame \cite{GlNe07}:
\begin{equation}
|R;IJ^{PC}\rangle=\sum_{L}\sum_{\lambda_q\lambda_{\qbar}}
\chi_{\lambda_q \lambda_{\qbar}}^{RPI}\\
\label{eq_ls_basis}
\sqrt{\frac{2L+1}{2J+1}}
C_{\frac12\lambda _q\frac12-\lambda_{\qbar}}^{S\Lambda}
C_{L0S\Lambda}^{J\Lambda}|I;{}^{2S+1}L_J\rangle\;,
\end{equation}
where the summation is implied in helicities of the fermion $\lambda_q$ and
antifermion $\lambda_{\qbar}$  as well as  in the orbital angular momenta  $L$
such  that $(-1)^{L+1}=P$. The total spin $S$ is fixed by the quantum numbers
$I J^{PC}$. Coefficients $\chi_{\lambda_q \lambda_{\qbar}}^{RPI}$ can be
extracted from Table~2 of Ref.~\cite{GlNe07}. It follows immediately from
Eq.~(\ref{eq_ls_basis}) that  every state (interpolator) in the chiral basis
is a fixed (prescribed by chiral  symmetry and unitarity) superposition of
allowed states in the $\{I;{}^{2S+1}L_J\}$  basis. For instance, there are two
kinds of the vector states (interpolators) with the quantum numbers of the
$\rho$-meson, which are represented by two orthogonal fixed combinations of
$S$- and $D$-waves:
\begin{equation}\label{unitary_1}
\left(
\begin{array}{l}
|(0,1)\oplus(1,0);1 ~ 1^{--}\rangle\cr
|(1/2,1/2)_b;1 ~ 1^{--}\rangle
\end{array}
\right) = U\cdot
\left(
\begin{array}{l}
|1;{}^3S_1\rangle\cr
|1;{}^3D_1\rangle
\end{array}
\right)
\end{equation}
with 
\begin{equation}\label{unitary_2}
U=
\left(
\begin{array}{cc}
\sqrt{\frac23} & \sqrt{\frac13} \cr
\sqrt{\frac13} & -\sqrt{\frac23} 
\end{array}
\right)\;.
\end{equation}
In terms of the quark-antiquark bilinears the state $|(0,1)\oplus(1,0);1 ~
1^{--}\rangle$ is given by the spatial components of the standard vector
current, $O_\rho^V$ from (\ref{rV}),  while the pseudotensor interpolator 
$O_\rho^T$  in (\ref{rT}) represents $|(1/2,1/2)_b;1 ~ 1^{--}\rangle$. 
\footnote{It is actually possible to construct interpolators with derivatives
that also transform according to the representations above \cite{Gl03,Gl07}.}
Consequently, diagonalizing the cross-correlation matrix with the $O_\rho^V$ and
$O_\rho^T$ interpolators and using the unitary  transformation we can
reconstruct the partial wave content of the $\qbar q$ component of the
$\rho$-meson.

\section{The variational method and the ``wave function'' of a hadron}
\label{sec_var_meth}

The composition of hadronic states in quantum field theory is a subtle issue. 
Whereas in non-relativistic approaches the notion of a wave function and a
complete  basis of states is well-defined, in QFT beyond the ground state
there is no well-defined single hadron, the state is always a scattering state
with superposition of many particle components. A given hadron interpolator
couples in principle to all states with its quantum numbers. 

In the lattice formulations this situation is further complicated by the
coarseness of the lattice and its symmetries, which mostly have no one-to-one
correspondence to the continuous space-time symmetries. Lattice interpolators
for mesons, in their simplest disguise, are  point like or extended color
singlet quark-antiquark  combinations with suitable combination of their Dirac
components. 

Combinations of such interpolators may then be formed such as to represent
lattice analogues of spin representations \cite{Li07a,GaGlLa08}.  Usually the
even or odd spin representations mix among them, thus, e.g., a lattice
interpolator representing a $J=1$ vector will have contributions which
correspond to higher odd $J$ states and thus corresponding signals in the
correlation function. Suitable combinations may improve the situation, but
much higher statistics is then necessary. This is one of the reasons why
lattice studies of excited states are progressing slowly.

The variational method \cite{Mi85,LuWo90}  (for a recent, more complete set of
references cf. \cite{BlDeHi09}) provides a method to find the optimal
combinations. Starting with a set of interpolators $\{O_1,O_2,\ldots,O_N\}$
one determines the correlation function (\ref{eq_corrfctn}) and solves a
generalized eigenvalue problem (see below).  Given a large enough basis set 
of interpolators the eigenvalues may then be related to the eigenstates of the
Hamiltonian and the corresponding energy values.

In order to avoid possible prejudices one should provide a large  basis of
interpolators. On the other hand, using too many such interpolators may
increase the statistical noise in the analysis. Since the optimal combination
models the hadron structure it is helpful to be guided by intuition -- which
is, however, another word for prejudice. Thus the selection of a suitable
basis is close to being an art and one has to balance quality with quantity.

The normalized physical states $|n\rangle$ propagate in time with
\begin{equation}
\langle n(t)|m(0)\rangle =\delta_{nm}\mathrm{e}^{-E^{(n)} t}\:.
\end{equation}
The interpolating (lattice) operators $O_i(t)$ are projected to vanishing
spatial momentum and are usually not normalized. 

We compute the correlation function
\begin{equation}\label{corr_inf}
C(t)_{ij}=\langle O_i(t)O_j^\dagger(0)\rangle=\sum_n^\infty a_i^{(n)} a_j^{(n)*} 
\mathrm{e}^{-E^{(n)} t}\;,
\end{equation}
with the coefficients giving the overlap of the lattice operator with the
physical state,
\begin{equation}\label{eq_w_f}
a_i^{(n)}=\langle 0| O_i|n\rangle\;.
\end{equation}
For interpolating operators $O_i$ spanning an orthogonal basis these values
would indeed constitute the wave function of  state $|n\rangle$ in that
basis (more exactly, the Bethe-Salpeter amplitude).
In this work we are using the term ``wave function'' for this matrix eleemnt..

We assume that the correlation matrix (\ref{corr_inf}) can be approximated by a
finite sum over  $N$ states and denote this  approximation by 
\be
\label{corr_finite}
\widehat C(t)_{ij}=\sum_{n=1}^N a_i^{(n)} a_j^{(n)*} \mathrm{e}^{-E^{(n)} t}\;.
\ee
The eigenvector and eigenvalues of $\widehat C$ will not exactly agree with those of $C$.

It can be shown \cite{Mi85,LuWo90} (for a recent discussion see 
\cite{BuHaLa08,BlDeHi09}) that the generalized eigenvalue problem (summation convention)
\begin{equation}\label{gev_1}
\widehat C(t)_{ij} u_j^{(n)} =\lambda^{(n)}(t,t_0)\widehat C(t_0)_{ij} u_j^{(n)}
\end{equation}
allows to recover the correct eigensystem approximately, i.e.,
\begin{equation}\label{gev_2}
\lambda^{(n)}(t,t_0)=\mathrm{e}^{-E^{(n)} (t-t_0)} 
\left(1+\mathcal{O}\left(\mathrm{e}^{-\Delta E^{(n)} (t-t_0)}\right)\right)
\;.
\end{equation}
Here $\Delta E^{(n)}$ may be as small as the distance to the next nearby energy
level. In \cite{BlDeHi09} it was pointed out that in an interval $t_0\le t \le
2\,t_0$ these contributions are suppressed and leading terms even have $\Delta
E^{(n)}$ equal to the distance to the first neglected energy level  $E^{(N+1)}$.
At $t_0$ all eigenvalues  are 1 and the eigenvectors are arbitrary.

Inserting (\ref{corr_finite}) into (\ref{gev_1}) one sees that the
eigenvectors of the generalized eigenvalue problem come out orthogonal to the
original wave functions $a^{(n)}$,
\begin{equation}\label{gev_dual}
(u^{(n)},a^{(m)})\equiv\sum_{i=1}^N u_i^{(n)*} a_i^{(m)}= c^{(m)}\delta_{nm}\;,
\end{equation}
and approximate the correct ones. The normalizing factor $c^{(m)}$  we can get
rid off.

Defining a sum of lattice operators
\begin{equation}
\eta^{(n)} \equiv \sum_{i=1}^N u_i^{(n)*} O_i\;,
\end{equation}
we find
\begin{equation}
\langle m|\eta^{(n)}|0\rangle =
\sum_{i=1}^N u_i^{(n)*} \langle 0|O_i^\dagger|m\rangle =
\sum_{i=1}^N u_i^{(n)*} a_i^{(m)}=c^{(n)}\delta_{nm}\;.
\end{equation}
Therefore $\eta$ creates a physical state,
\begin{equation} 
\eta^{(n)\dagger}|0\rangle=c^{(n)*}|n\rangle\;.
\end{equation}
The eigenvector coefficients give the  composition of the eigenstate in terms
of the (non-orthogonal)  interpolating operators:
\begin{equation}
a_i^{(n)}=\langle 0|O_i|n\rangle
=\frac{1}{c^{(n)*}}\langle 0|O_i\,\eta^{(n)\dagger}|0\rangle
=\frac{1}{c^{(n)*}}\sum_{j=1}^N u_j^{(n)} \langle 0| O_i O_j^\dagger|0\rangle\;.
\end{equation}
They would agree with $u_j^{(n)}$ if the interpolators were orthogonal, which
they are usually not. 

The eigenvectors of the generalized eigenvalue problem for hermitian matrices
$\widehat C(t)$, $\widehat C(t_0)$ obey the orthogonality relation
\begin{equation}
\left(u^{(n)},\widehat C(t)u^{(n)}\right)\propto \delta_{nm}\;. 
\end{equation}
Indeed, with (\ref{corr_inf}) and (\ref{gev_dual})
we find for large $t$ (summation convention)
\begin{eqnarray}
w_i^{(n)}(t)\equiv \widehat C(t)_{ij} u_j^{(n)}
&\sim &c^{(n)*} a_i^{(n)} \mathrm{e}^{-E^{(n)} t}\;,\nonumber\\
\left(u^{(n)},w^{(n)}\right)=
u_i^{(n)*} \widehat C(t)_{ij} u_j^{(n)}&=& c^{(n)*} u_i^{(n)*} a_i^{(n)}
\mathrm{e}^{-E^{(n)} t}\nonumber\\
&=&\left|c^{(n)}\right|^2 \mathrm{e}^{-E^{(n)} t}\;,
\end{eqnarray}
and for the ratio
\begin{equation}
\frac{\left|w_i^{(n)}\right|^2}{(u^{(n)},w^{(n)})}
= \left|a_i^{(n)}\right|^2 \mathrm{e}^{-E^{(n)}t}\;.
\end{equation}
Assuming asymptotically leading exponential behavior this allows to read off
$|a_i^{(n)}|$ in the asymptotic region. Ideally $a_i^{(n)}$ should not depend
on $t$, in actual calculations, however, one identifies a region of $t$-value
where it is compatible with a constant.

We may utilize this result to determine ratios of couplings of  the different
lattice operators to the physical states,
\begin{equation}\label{ratio_op_comp}
\frac{w_i^{(n)}(t)}{w_k^{(n)}(t)}=\frac{\widehat C(t)_{ij} u_j^{(n)}}{\widehat C(t)_{kj} u_j^{(n)}}=\frac{a_i^{(n)}}{a_k^{(n)}}\;.
\end{equation}
One computes the ratio for several values of $t$ and identifies its value in a
plateau region. The ratio tells us how much different interpolating operators
contribute to the eigenstate $|n\rangle$. This can be used to discuss
contributions of, e.g., different representations of the vector meson channel,
as we do here.

In a realistic simulation many physical states may contribute and one has to
use the discussed techniques to single out couplings to the state of interest.
In some approximation one may instead look just at the ratios of individual
entries of the correlation matrix, like
\be\label{ratio_corr_entries}
\widehat C_{ii}(t)/\widehat C_{jj}(t) \quad\textrm{or}\quad 
\widehat C_{ii}(t)/\widehat C_{ij}(t) 
\ee
in the asymptotic $t$-region, where the excited states contributions are
suppressed \cite{BeLuMe03,BrBuGa03,AlAnAo08}.

The ratio
\begin{equation}
\left|\frac{a_i^{(n)}}{a_i^{(m)}}\right|^2=
\left|\frac{\langle 0|O_i(t)|n\rangle}{\langle 0|O_i(t)|m\rangle}\right|^2
\end{equation}
tells us how much the interpolating operator $O_i$ contributes to the
eigenstates $|n\rangle$ and $|m\rangle$. This can be used to discuss the ratio
of decay constants of various excitations as done in 
\cite{BuEh07,BuHaLa08,BlDeHi09}.

\section{Vector meson couplings for local interpolators} \label{sec_veccoup}

Here we quote definitions of the local coupling constants of the vector and
pseudotensor currents to the $\rho$-meson in continuum and discuss their
relations to the matrix elements obtained in the previous section.

In Minkowski space the corresponding amplitudes are given as
\begin{eqnarray}
 \langle 0 | \qbar(0) \gamma^\mu q(0) | V(p; \lambda)\rangle &=& 
 m_\rho f_\rho^V e^\mu_\lambda\;,
\label{rhoV}\\
 \langle 0| \left(\qbar(0) \sigma^{\alpha \beta} q(0)\right)(\mu) | V(p; \lambda)\rangle &= &
 \I f_\rho^T(\mu) e^\mu_\lambda (e^\alpha_\lambda p^\beta -  e^\beta_\lambda p^\alpha)\;.
\label{rhoT}
\end{eqnarray}
Here   $V(p; \lambda)$ is the vector meson state with the mass $m_\rho$,
momentum $p$ and polarization $\lambda$. The vector current is conserved,
consequently the vector coupling constant $f_\rho^V$ is scale-independent. The
pseudotensor ``current'' is not conserved and is subject to a nonzero
anomalous dimension. Consequently the pseudotensor coupling $ f_\rho^T(\mu)$
manifestly depends on the  scale $\mu$. In the rest frame the ratio
\begin{equation} 
\frac{f_\rho^V}{f_\rho^T(\mu)} =  \frac
 {\langle 0 | \qbar(0) \gamma^i q(0) | V(\lambda)\rangle}
 {\langle 0 |\left(\qbar(0) \sigma^{0i} q(0)\right)(\mu) | V(\lambda)\rangle}
 \label{rhoV/rhoT}
\end{equation}
coincides with the ratio of matrix elements (\ref{ratio_op_comp}).

That ratio can be extracted from the  ratio of the vector-vector and
pseudotensor-vector   zero-momentum correlators at large Euclidean times, when
the excited  states do not contribute any more. If the source is located at
the point $(t,\vec x) = (0,\vec 0)$ and the sink is at the arbitrary point
$(t,\vec x)$, then at large $t$ one obtains asymptotically
\begin{eqnarray}
 \sum_{\vec x} \langle 0 |\left(\qbar(t,\vec x) \gamma^i q(t,\vec x) \right) 
 \left(\qbar(0) \gamma^i q(0)\right)^\dag | 0 \rangle &\sim& (f_\rho^V)^2 m_\rho^2 \exp(-m_\rho t)\;,
\label{VV}\\
\sum_{\vec x} \langle 0 | \left(\qbar(t,\vec x) \sigma^{0i} q(t,\vec x)\right)(\mu)  
 \left(\qbar(0) \gamma^i q(0)\right)^\dag | 0 \rangle &\sim& 
 f_\rho^V f_\rho^T(\mu) m_\rho^2 \exp(-m_\rho t)\;.
\label{VT}
\end{eqnarray}
The ratio of these correlators is equal to the ratio $f_\rho^V/
f_\rho^T(\mu)$. With the variational method the ground state can be identified
already at small  Euclidean time distance and thus the quality of the result
can be improved, see Sect.\ \ref{Sim_details}.

In \cite{BeLuMe03,BrBuGa03,AlAnAo08} this ratio has been used to extract
$f_\rho^V/ f_\rho^T(\mu)$ and to relate\footnote{Here there is a subtle point.
In the \MSbar-scheme of continuum theory one uses a renormalization group
equation obtained to typically two or three loops to relate the ratio at
different scales. Such a renormalization group equation cannot adequately
represent the physics related to chiral symmetry breaking, because chiral
symmetry breaking is intrinsically a nonperturbative effect. If, at given
scale, the $\rho$-meson couples strongly to  both, vector and pseudotensor
interpolators, then chiral symmetry is strongly broken in the $\rho$-meson at
that scale.  If chiral symmetry is broken, however, it is unclear whether
perturbative RG is applicable to relate the ratio $f_\rho^V/ f_\rho^T(\mu)$ at
different scales.}  it to the continuum \MSbar-scheme at a scale of 2 GeV.
While this method is suitable to study the ratio for the ground state and the
result agrees with that extracted from the variational method (to be compared
later on), only the variational method can be used to study such ratios for
excited states.

\section{Smearing of the interpolators and the resolution scale}\label{sec_smearing}

So far we discussed the cross-correlation matrix as obtained with the local
interpolators. In lattice simulations the hadron interpolators can be built
with different spatial extent, e.g., using so-called smeared quark sources. 
For example, an isovector interpolator  may have the form 
\begin{equation}
\dbar(x')\,S_{x'0}^\dagger\,\Gamma\, S_{0x''}\,u(x'')\;,
\end{equation}
where we omit the Dirac indices; $\Gamma$ denotes some Dirac matrix and 
$S_{xy}$ is some gauge transporter from $y$ to $x$.  For the local
interpolator at the origin one has $x' = x'' = 0$. Summation over $x'$ and
$x''$ may be used to define quark source smearing, thus improving the signal
when computing correlators of such interpolators.

One purpose of such a smearing is to improve the quality of the signal from
the state of interest in order to extract, e.g., its mass. Indeed, with the 
local interpolator the coupling of this interpolator to the physical state is
determined by the behavior of this state at the origin. With the local
interpolators we therefore probe the hadron wave function at the scale fixed
by the lattice  spacing $a$. If we smear the local interpolator in a
gauge-invariant way over a spatial region of size $R$, then the coupling of
our interpolator to the physical state may be better and the quality of the
signal is improved. So the smearing typically plays a technical role. In our
case, however, we give the smearing width a fundamental role -- it defines a
resolution scale at which we study the hadron wave function. Of course, different smearing
methods may lead to different definitions.
 
Only a few quantities in QCD do not depend on the scale. For example, the
lattice spacing $a$ fixes the ultraviolet cut-off (i.e., the renormalization
scale), and the observables such as the ratios of masses of different hadrons,
their electric charges  should not depend on this scale. Most of the
quantities in QCD do depend on the resolution scale at which they are studied.
We want to study the hadron wave function in the infrared (where mass is
generated), i.e., at the very low resolution scale characterized by the
typical hadron size. Certainly we cannot chose $a$ to be so large since then
we lose  matching to the ultraviolet (continuum) limit of QCD. However, even
if we use a reasonably small $a$ we can fix a scale where we study the hadron
at the smearing size $R$. If $R \gg a$, it is the size
of the smearing $R$ that defines a resolution scale where we probe the hadron
properties. Consequently, the smearing plays a  rather fundamental role --
it defines a scale at which we study the content of our hadron. Physically it
means that given a source smearing size $R$, we cannot resolve details of our
wave function with the smaller size. Physical (continuum) results can be
deduced from the extrapolation to the $a = 0$ point while keeping $R$ fixed
in physical units.

We fix the resolution scale in the following way. We substitute a local
interpolator at the point $(t,\vec x)$ by the interpolator with the same
quantum numbers but with quark fields smeared in spatial coordinates over the
size $R$ in a  gauge-invariant way around the point $(t,\vec x)$. The profile
of the smeared quark fields should be approximately Gaussian with the width
$R$. Therefore we use the so-called Jacobi smearing \cite{Gu89,Be97}. A
point-like source $S_0$ is smeared out by acting with a smearing operator $M$,
\be
S = M S_0\ , \quad M=\sum_{n=0}^N\, (\kappa H)^n\ ,
\ee
where $H$ is a hopping term,
\be
H=\sum_{j=1}^3\, \left[ U_j(\vec{x},\,t)\,\delta_{\vec{x}+\hat{j},\vec{y}} +
U_j^\dagger(\vec{x}-\hat{j},\,t)\,\delta_{\vec{x}-\hat{j},\vec{y}} \right]\;.
\ee
The smearing extends only over individual time slices, i.e., $t$ is fixed. The
parameters $\kappa$ (hopping parameter) and $N$ (number of smearing steps) are
tuned to get an approximately Gaussian shape of the quark source with a
certain width $R$ in physical units. Different smearing algorithms may
be used for different definitions of the resolution scale.

\section{Simulation details and results}\label{Sim_details}

In earlier work we have studied so-called chirally improved (CI) fermions
\cite{Ga01a,GaHiLa00} in the quenched 
\cite{GaGoHa03a,BuGaGl04a,BuGaGl06,GaGlLa08} and the dynamical \cite{GaHaLa08}
case for two mass-degenerate light quarks. The gauge field action was the
tadpole improved L\"uscher-Weisz action \cite{LuWe85}.  Table \ref{tab_data}
gives some information on the runs analyzed here, details on the simulation
and other observables (e.g., hadron masses) can be found in the original
papers.  The spatial lattice size 16 corresponds to a physical size close to
2.4 fm, the temporal size is twice as large.

In the quenched simulations the $\rho$ cannot decay; in our dynamical
ensembles \cite{GaHaLa08} its mass is also above the decay threshold since
extra units of relative (quantized lattice) momentum are needed for  the
decay. Below threshold the 2-pion states will become important and probably
modify our results. However, we study here only the quark-antiquark
contributions to the $\rho$ lattice state. Adding further, more-quark
interpolators may affect the overall normalization of the quark-antiquark
component but is unlikely to change the ratios of the vector vs.
pseudovector operators of the quark-antiquark contribution. Only this ratio
is important for the partial wave decomposition.

\begin{table}[tb]
\caption[]{
Specification of the data used here; for the gauge coupling only the leading
value $\beta_{LW}$ is given, $m_0$ denotes the bare mass parameter of the
CI-action. Further details like  the determination of the lattice spacing and
the $\pi$- and $\rho$-masses are found in \cite{GaGlLa08,GaHaLa08}. For the
quenched case and for the dynamical ensemble A we used 100 configurations, for
sets B and C we analyzed 200 configurations each. The lattice size is
$16^3\times 32$ throughout.}\label{tab_data}
\begin{center}
\begin{tabular}{lccccc}
\hline\noalign{\smallskip}
Data&  $\beta_{LW}$ &  $a\,m_0$& $a$ [fm] & $m_\pi$[MeV]& $m_\rho$[MeV]\\
\noalign{\smallskip}\hline\noalign{\smallskip}
Quenched& 7.90 & 0.04--0.20&   0.1480(10)   &475--1053     &  912--1251\\
dyn.: A & 4.70 & -0.050&       0.1507(17) &526(7)& 922(17)\\
dyn.: B & 4.65 & -0.060&       0.1500(12) &469(4)& 897(13)\\
dyn.: C & 4.58 & -0.077&       0.1440(12) &318(5)& 810(28)\\
\noalign{\smallskip}\hline
\end{tabular}
\end{center}
\end{table}

In our analysis we use the variational method, with quark-bilinear meson 
interpolators. We use Jacobi smearing of quark sources with the values for
$\kappa$ and $N$ such as to obtain a narrow (index $n$) and a wide (index $w$)
source with effective smearings widths \cite{BuGaGl04a} of 0.27 fm and 0.41 fm for the quenched
ensembles and 0.27 fm and 0.55 fm for the dynamical ones, respectively
\cite{GaGlLa08,GaHaLa08}.

The variational method not only allows to identify the excited state(s) --
depending on the number of interpolators used and the statistical quality of
the data -- but also gives increased stability in the ground state signal.
Here we discuss only results obtained for the ground state in the isovector
vector channel, the $\rho$-meson ($J^{PC}=1^{--}$). We include the operators
\begin{eqnarray}
O^V_n=\overline u_n \gamma^i d_n\;,\;\;
&O^V_w=\overline u_w \gamma^i d_w\;,\;\;
&O^V_p=\overline u_p \gamma^i  d_p\;,\;\;\nonumber\\
O^T_n=\overline u_n \gamma^t \gamma^i  d_n\;,\;\;
&O^T_w=\overline u_w \gamma^t \gamma^i  d_w\;,
&O^T_p=\overline u_p \gamma^t \gamma^i  d_p\;,
\end{eqnarray}
where $\gamma^i$ is one of the spatial Dirac matrices, $\gamma_t$ is the
$\gamma$-matrix in (Euclidean) time direction, and the subscripts $w$ and $n$ denote
the two smearing widths of the quark sources, whereas $p$ indicates point quark
sources. We denote the ratios of the coupling of the $\rho$-meson to the different
interpolators by $[a_\rho^V/a_\rho^T](n)$ for the narrow quark sources and 
analogously for the other two cases $w$ and $p$.

\begin{figure}[tbp]
\begin{center}
\includegraphics*[width=10cm]{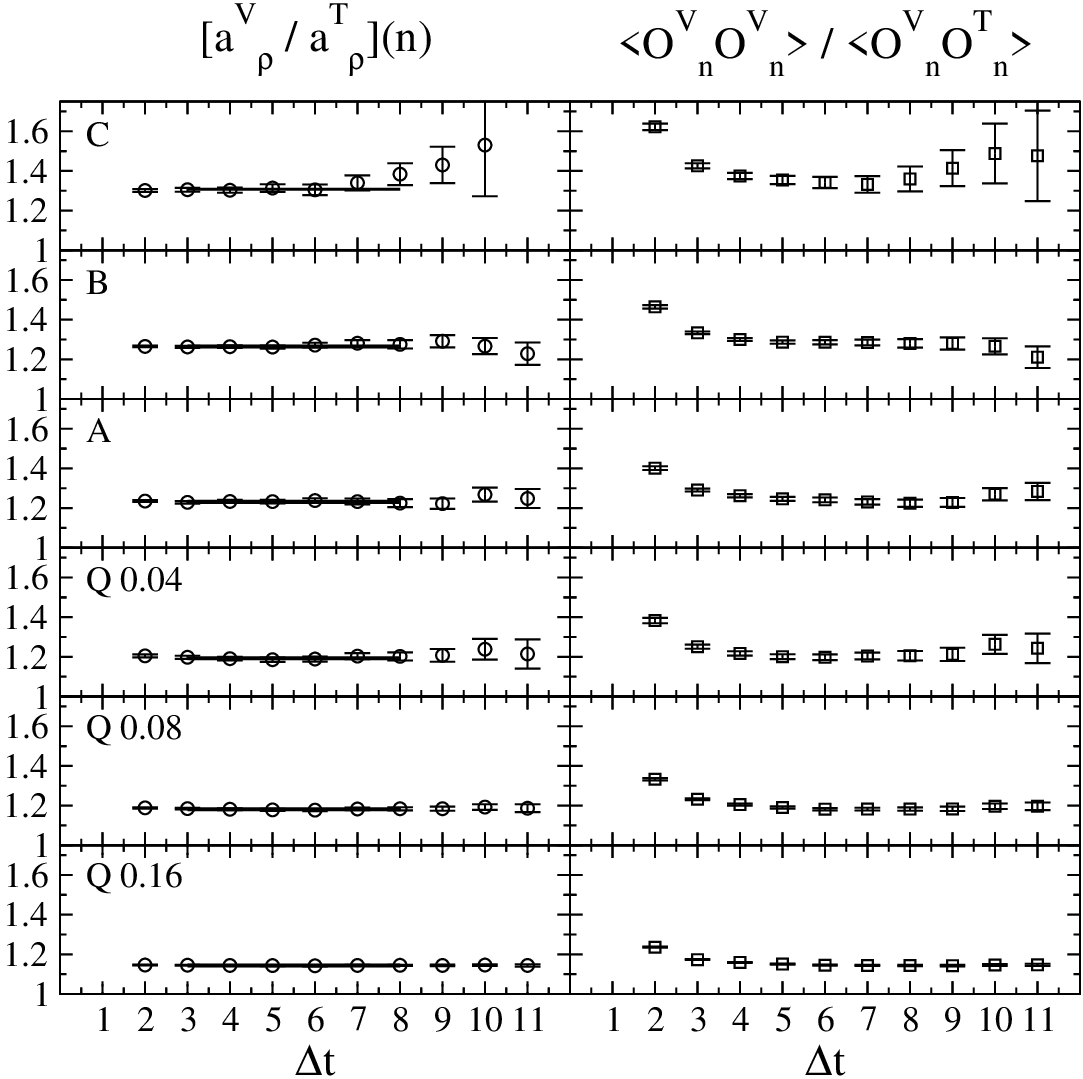}
\end{center}
\caption{\label{fig:plot_2_6}
We compare for $a_\rho^V/a_\rho^T$ the a-ratios (left-hand colum)
determination with the d-ratio (right-hand column)
determination for dynamical and some quenched data sets.
The error bars have been determined with single-elimination jack-knife.
The horizontal lines indicate the plateau fit range and value used in Fig. \ref{fig:plot_ratios_mass}.
The data correspond to ensembles given in Table \ref{tab_data}, for the quenched ensemble
we show only the results for valence masses 0.04, 0.08 and 0.16 (in lattice units).}
\end{figure}

In Fig.\ \ref{fig:plot_2_6} we compare the ratios (which we call a-ratios henceforth)
of the coefficients determined from diagonalization of the correlation matrix between
interpolators  $O^V_n$, $O^V_w$, $O^T_n$, $O^T_w$ as discussed in \eq{ratio_op_comp}
with the direct ratio between matrix elements (called d-ratios) according to
\eq{ratio_corr_entries}.

The plateaus of the a-ratios are remarkable wide and stable for large quark masses and
decrease in quality towards smaller masses. However, even for run C with $m_\pi$ close
to 320 MeV  we still observe good quality plateaus for time separations  $\Delta
t=3\ldots 8$. The direct d-ratios  have larger errors and much worse plateau
behavior;  for run C only the range $\Delta t=5\ldots 8$ is acceptable. The
contamination with excited states at smaller distances is obvious. The values are,
however, compatible with the a-ratios. We also confirm that, e.g., the d-ratios 
$\langle O^V_n O^V_n\rangle/\langle O^T_n O^T_n\rangle$
are approximately
equal to 
$\left(\langle O^V_n O^V_n\rangle/\langle O^V_n O^T_n\rangle\right)^2$.

The data for the wide sources are qualitatively similar. We therefore use the a-ratio averages
of the plateau values from 3 to 8 as our value estimate for $[a_\rho^V/a_\rho^T](n)$ for the
narrow sources and $[a_\rho^V/a_\rho^T](w)$ for the  wide sources.

The quark-propagators for point sources were not available but we do have correlation matrix
entries between smeared source interpolators and point sink interpolators.  Due to that
limitation we cannot use the variational method (we do not have the full correlation matrix) but
determine the results via d-ratios of entries of the correlation matrix giving
$[a_\rho^V/a_\rho^T](p)$.

\begin{figure}[tp]
\begin{center}
\includegraphics*[width=7cm]{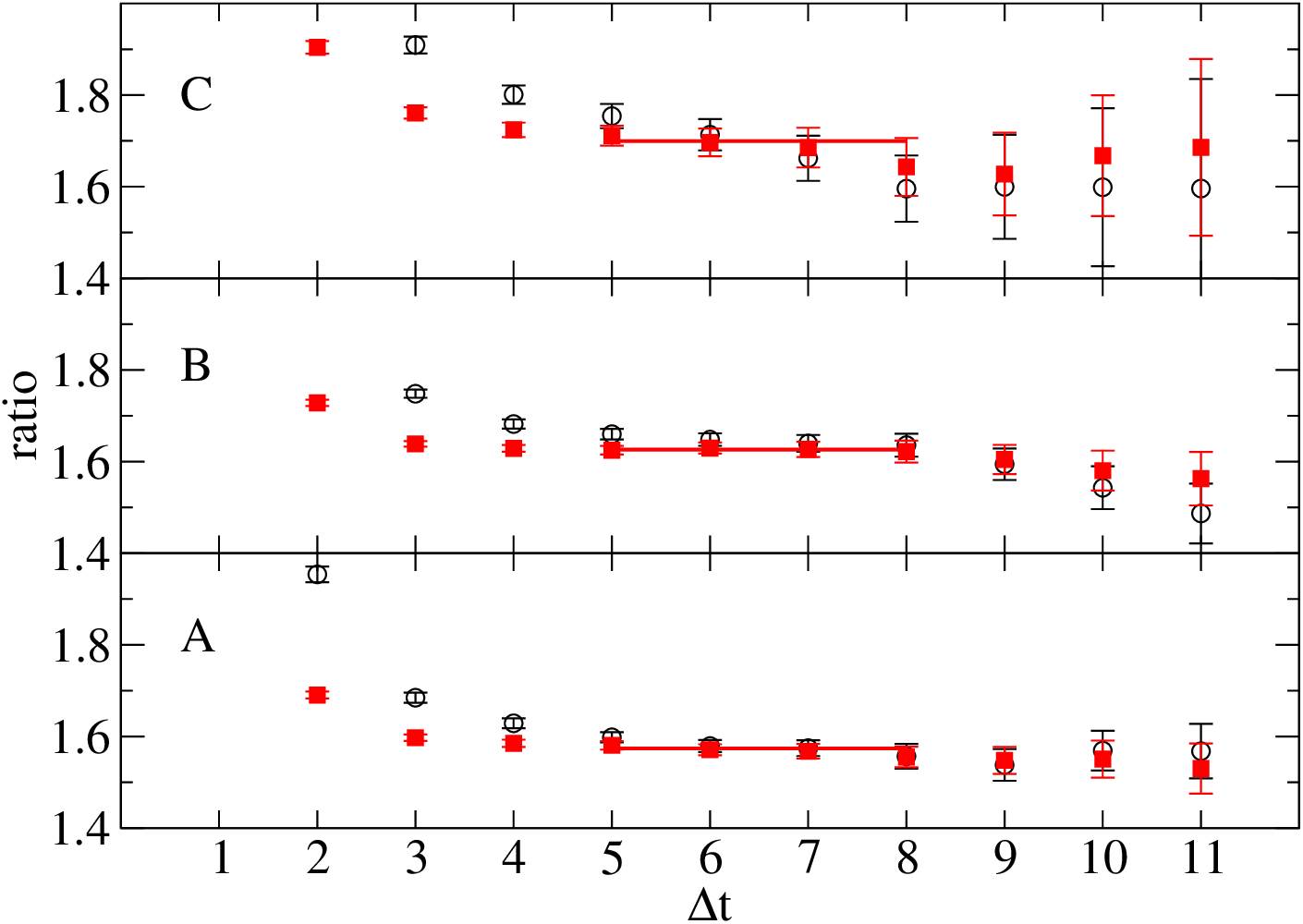}
\end{center}
\caption{\label{fig:plot_1_3}
We compare (for the dynamical data runs A, B and C) ratios of correlation matrix
entries (see \eq{rat_n_w_p}), which in the plateau region should give
$[a_\rho^V/a_\rho^T](p)$.  The results for narrow (open symbols) and wide (full
symbols) sources agree in the plateau region, but the plateau range shrinks and is
hardly justifiable for run C. The error bars have been determined with
single-elimination jack-knife. The horizontal lines indicate the plateau fit range
(for the $w$ sources) and the fit value used in Fig. \ref{fig:plot_ratios_mass}.}
\end{figure}

\begin{figure}[tp]
\begin{center}
\includegraphics*[width=7cm]{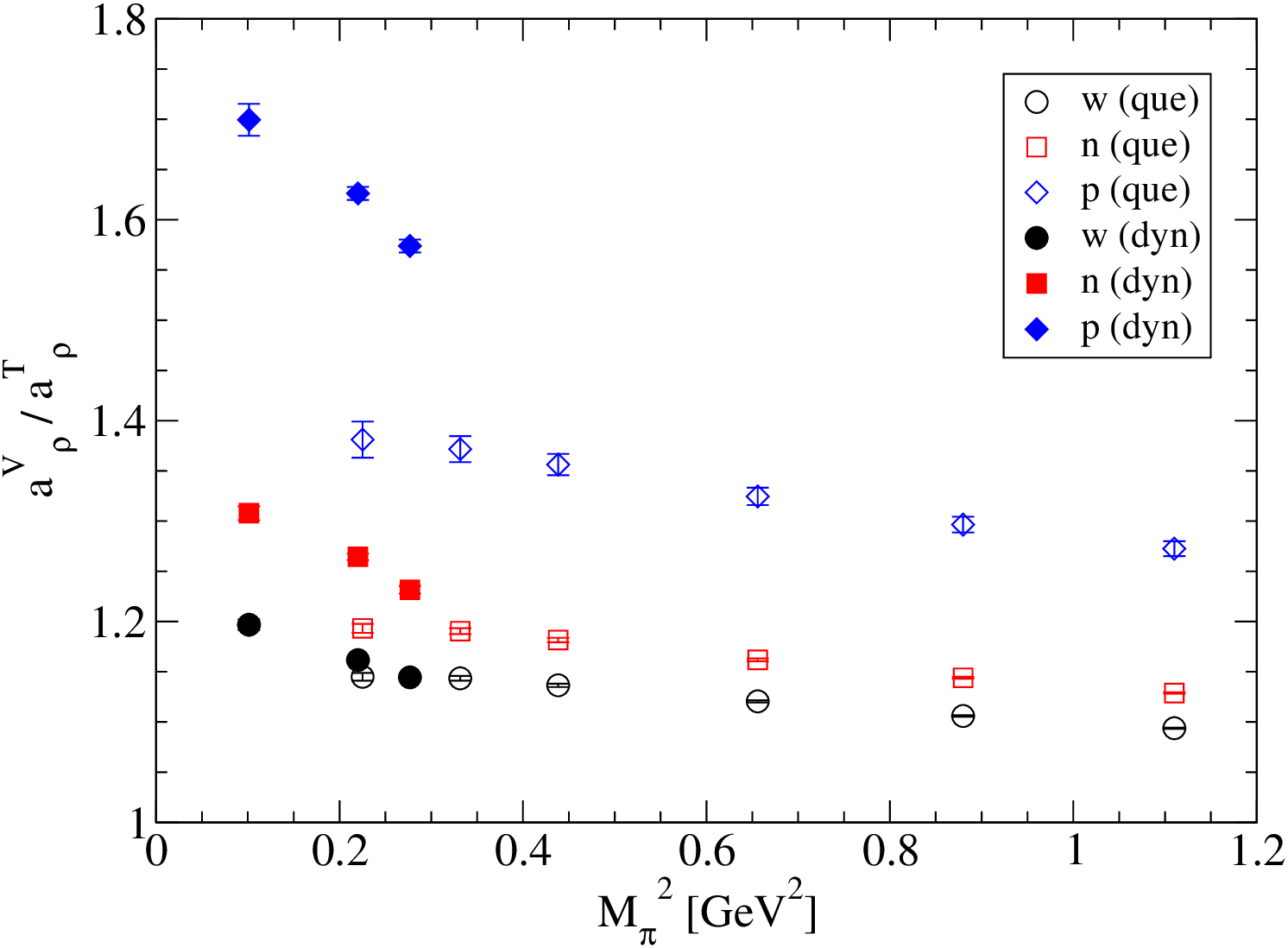}
\end{center}
\caption{\label{fig:plot_ratios_mass}
We compare our results for $[a_\rho^V/a_\rho^T]$  for the three smearing scales: $w$,
$n$ and $p$ (no smearing). The open symbols denote quenched data, the full symbols the
data with dynamical quarks. The error bars have been determined with
single-elimination jack-knife.}
\end{figure}

In the ratios 
\be\label{rat_n_w_p}
\left(\langle O^V_nO^V_p\rangle/\langle O^V_nO^T_p\rangle\right)^2\quad\textrm{and}
\quad
\left(\langle O^V_wO^V_p\rangle/\langle O^V_wO^T_p\rangle\right)^2
\ee
ideally the effect of the smeared sources should cancel and the ratios should agree.
In Fig. \ref{fig:plot_1_3} we compare them and find agreement in the plateau region, 
which, however is shrinking and hardly identifiable for the run C with smallest quark
mass. It is interesting to note that the wide source ratios appear to have less
contamination from excited states. We use these for the values in  Fig.
\ref{fig:plot_ratios_mass}.

Fig. \ref{fig:plot_ratios_mass} exhibits our results for $[a_\rho^V/a_\rho^T]$ for all
three situations of interpolator smearing: $w$, $n$ and $p$ comparing quenched with
dynamical data. A systematic dependence on the smearing scale of the interpolators is
obvious.  We also find that towards smaller quark masses the quenched results (i.e.,
for the lowest valence masses 0.04 and 0.06 in lattice units) significantly deviate
from the data with dynamical quarks.

We  observe a clear and systematic dependence of $a_\rho^V/a_\rho^T$ on the smearing
properties of the hadron interpolators, more precisely:  the quark sources building
the interpolators. As discussed in Sect.\ \ref{sec_smearing} the dependence on the
smearing width (ranging from 0.55 fm down to the unsmeared point scale $a \approx
0.15$ fm) allows to relate the composition of the $\rho$ on various infrared
resolution scales.  A more systematic study of that scale dependence may allow  to
better compare with effective models. 

Physically the ratio encodes how the chiral symmetry is broken in the (quark-antiquark
components of the) $\rho$ wave
function at different scales. The $(0,1)\oplus(1,0)$ (i.e., the vector) and the
$(\frac{1}{2},\frac{1}{2})_b$ (i.e., the pseudotensor)  representations are a complete
set for the $\qbar q$ component of the $\rho$-meson. Hence, chiral symmetry is broken
in the $\rho$ wave function such that the $\qbar q$ component is a superposition of
both with  a relative ratio shown in Fig. \ref{fig:plot_ratios_mass}. There we see
clear evidence for the resolution scale dependence of chiral symmetry breaking. 
Extrapolating to
physical quark masses (where we have to assume that opening the 2-pion decay channel
would not affect the result significantly)
we expect that the ratio varies from $\approx 1.75$ for point
interpolators (a resolution scale given by the lattice spacing $a\approx 0.15$ fm)
down to  $\approx 1.25$ for the interpolators built with smeared, wide quark sources,
where the  resolution scale is given from the smearing width $R\approx 0.55$ fm.

Inverting the unitary transformations \eq{unitary_1}--\eq{unitary_2} we conclude that
the $\qbar q$ component of the $\rho$-meson is varying from predominantly $^3S_1$ wave
for the large $R$ values with increasing admixture of $^3D_1$ towards smaller $R$,
e.g., 
\begin{eqnarray} 
\frac{a_\rho^V}{a_\rho^T}=1.75 &\rightarrow& 0.995|1;{}^3S_1\rangle+0.096|1;{}^3D_1\rangle\;,\nonumber\\ 
\frac{a_\rho^V}{a_\rho^T}=\sqrt{2} &\rightarrow&|1;{}^3S_1\rangle\;,\nonumber\\ 
\frac{a_\rho^V}{a_\rho^T}=1.25 &\rightarrow& 0.998|1;{}^3S_1\rangle-0.059|1;{}^3D_1\rangle \;.
\end{eqnarray}

\begin{figure}[tb]
\begin{center}
\includegraphics*[width=8cm]{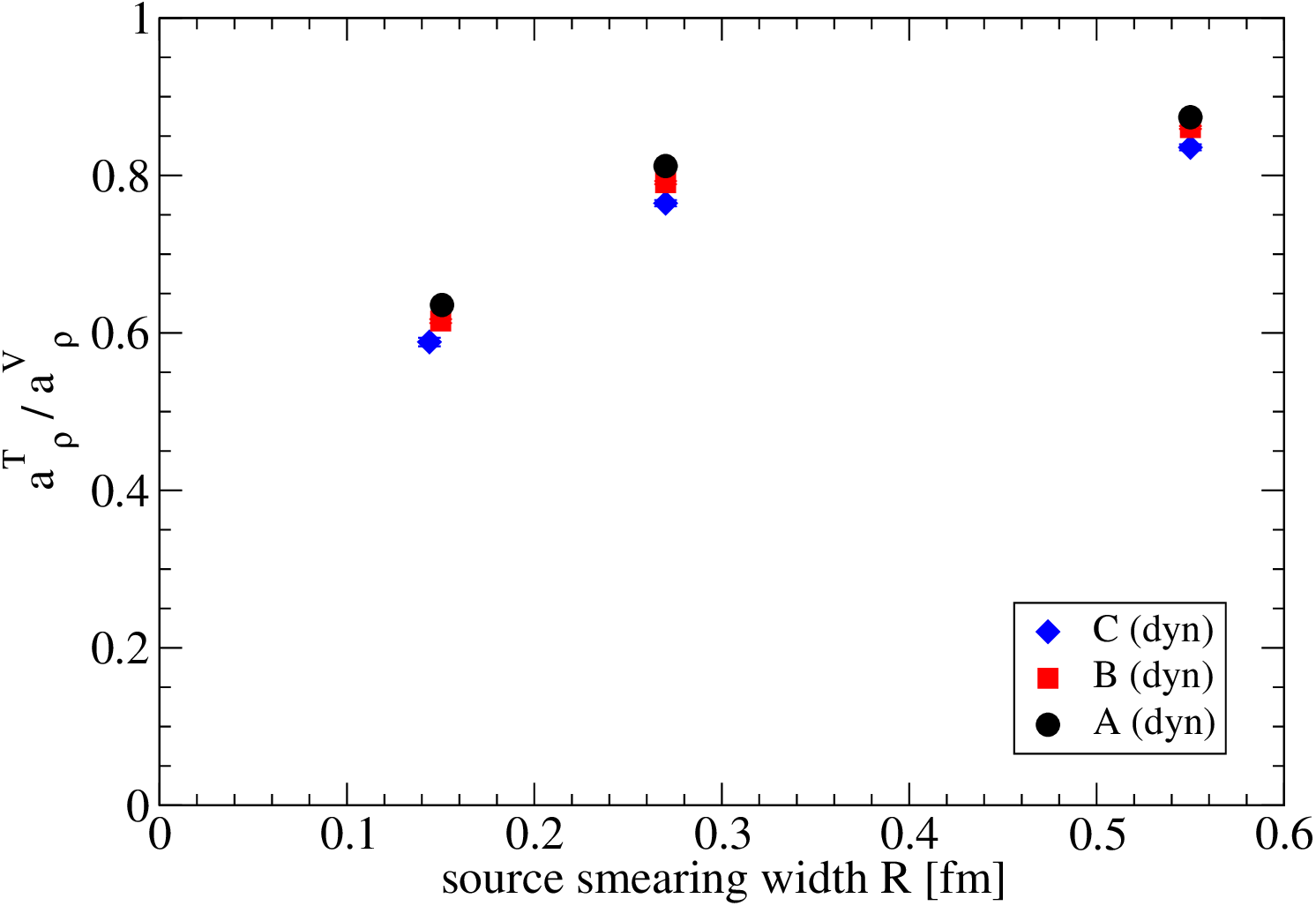}
\end{center}
\caption{\label{fig:IR_scale_dep}
We show the dependence of $[a_\rho^T/a_\rho^V]$ on the width of the quark
sources. For the point sources we use the lattice spacing $a$ as source width
$R$. Results for dynamical quarks:  A (circles), B (squares), C (diamonds).}
\end{figure}

In Fig. \ref{fig:IR_scale_dep} we show the inverse ratio $a_\rho^T/a_\rho^V$. 
Towards $\mu\sim 1/R\to\infty$ the pseudotensor contribution decouples from the
$\rho$-meson, $f_\rho^T(\mu\to\infty)=0$, as follows from renormalization group
behavior \cite{BeLuMe03} in the asymptotic freedom regime. Hence the inverse
ratio should approach 0 for $R\to 0$, fully compatible with Fig.
\ref{fig:IR_scale_dep}. At this point the partial wave decomposition is
determined by the $(0,1)\oplus(1,0)$ representation alone. For large $R$ the
inverse ratio saturates with the  $^3S_1$ partial wave strongly dominating over $^3D_1$,
with little dependence on the quark (or pion) mass.

\section{Conclusion}

In this paper we have defined and calculated the chiral and partial wave content of the $\qbar
q$ component of the $\rho$-meson at different resolution scales. We have used a complete (in
terms of chiral symmetry) basis of interpolators that allows to define the chiral symmetry
content of the quark-antiquark component of the $\rho$-meson. We have studied this in lattice
simulations in the quenched limit as well as for configuration obtained with $n_f = 2$
dynamical, mass-degenerate light quarks. We have computed the cross-correlation matrix of the
interpolators and applied the variational method for our analysis. The eigenvectors of the 
cross-correlation matrix supply us then with the direct information about decomposition of the
quark-antiquark component of the $\rho$-meson in terms of different representations of the
chiral group. 

Given such a decomposition we were able, using the unitary transformation from the chiral basis
to the $LSJ$-basis, to reconstruct the partial wave decomposition of the $\rho$-meson at
different resolution scales. It turns out that at low resolution scales $R \sim 0.2 - 0.6$ fm
the quark-antiquark component of the $\rho$-meson is a strong mixture of two representations of
the chiral group $(0,1)\oplus(1,0)$ and $(1/2,1/2)_b$ and consequently the chiral symmetry is
strongly broken at these infrared scales. Only at the deep ultraviolet scale asymptotic freedom
requires that the composition of the $\rho$-meson is be determined by the  $(0,1)\oplus(1,0)$
representation alone. Consequently, at low resolution $R \sim 0.2 - 0.6$ fm the $\rho$ wave
function is predominantly $^3S_1$ wave with a tiny admixture of the $^3D_1$ wave depending on
the scale. Only in the deep ultraviolet the $\rho$ is given by the fixed
$\sqrt{2/3}\,|I=1; ^3S_1\rangle + \sqrt{1/3}\,|I=1; ^3D_1\rangle$ superposition of the $S$- and
$D$-waves with a sizeable contribution of the $D$-wave. This explains successes of the $SU(6)$
flavor-spin-symmetry for the $\rho$-meson, that explicitly relies on the $^3S_1$ content of the
$\rho$ wave function. Note, however, that we observe a resolution
scale dependence of the composition, whereas the quark model does not.

\subsubsection*{Acknowledgement}
We thank G.\ Engel, C.\ Gattringer and D.\ Mohler for discussions. L.Ya.G.\ and M.L.
acknowledge support of the Fonds zur F\"orderung der Wissenschaflichen Forschung
(P19168-N16) and (DK W1203-N08), respectively. C.B.L. acknowledges support by 
DFG project SFB/TR-55.
The calculations have been performed on the SGI
Altix 4700 of the Leibniz-Rechenzentrum Munich and on
local clusters at ZID at the University of Graz.

\end{document}